\documentclass[a4paper,11pt]{article}

\usepackage{jheppub} 

\usepackage[T1]{fontenc} 

\newcommand{\slsh}{\rlap{$\;\!\!\not$}}     
\newcommand{\vH}{\mathbf{H}}
\newcommand{\vW}{\mathbf{W}}
\newcommand{\vB}{\mathbf{B}}
\newcommand{\vD}{\mathbf{D}}
\newcommand{\LL}{\mathcal{L}}
\newcommand{\tr}[1]{\operatorname{tr}\left[#1\right]}
\newcommand{\off}{\rm off}

\def\bentarrow{\:\raisebox{1.3ex}{\rlap{$\vert$}}\!\rightarrow}

\def\bothdk#1#2#3#4#5{
\begin{array*}{r c l}
#1 & \rightarrow & #2#3 \\
 & & \:\raisebox{1.3ex}{\rlap{$\vert$}}\raisebox{-0.5ex}{$\vert$}
\phantom{#2}\!\bentarrow #4 \nonumber \\
 & & \bentarrow #5
\end{array*}
}
\def\bentarrow{\:\raisebox{1.3ex}{\rlap{$\vert$}}\!\rightarrow}
\def\bothdk#1#2#3#4#5{
\begin{equation}
\begin{array}{r c l}
#1 & \rightarrow & #2#3 \\
 & & \:\raisebox{1.3ex}{\rlap{$\vert$}}\raisebox{-0.5ex}{$\vert$}%
\phantom{#2}\!\bentarrow #4 \\
 & & \bentarrow #5
\end{array}
\end{equation}
}

\def\kV{\kappa_V}

\def\beq{\begin{equation}}
\def\eeq{\end{equation}}
\def\beqn{\begin{eqnarray}}
\def\eeqn{\end{eqnarray}}

\title{Higgs constraints from vector boson fusion and scattering}

\author{John M. Campbell}

\author{and R. Keith Ellis}
\affiliation{Fermilab,\\PO Box 500, Batavia, IL 60510, USA}

\emailAdd{johnmc@fnal.gov}
\emailAdd{ellis@fnal.gov}

\abstract{

We present results on 4-lepton + 2-jet production, the partonic
processes most commonly described as vector boson pair production in
the Vector Boson Fusion (VBF) mode.  This final state contains
diagrams that are mediated by Higgs boson exchange.  We focus
particularly on the high-mass behaviour of the Higgs boson mediated diagrams,
which unlike on-shell production, gives information about the Higgs
couplings without assumptions on the Higgs boson total width.  We
assess the sensitivity of the high-mass region to Higgs coupling
strengths, considering all vector boson pair channels,
$W^-W^+$, $W^\pm W^\pm$, $W^\pm Z$ and $ZZ$.  Because of the small background,
the most promising mode is $W^+ W^+$ which has sensitivity to Higgs couplings
because of Higgs boson exchange in the $t$-channel.  Using the Caola-Melnikov
(CM) method, the off-shell couplings can be interpreted as bounds on the
Higgs boson total width. We estimate the bound that can be obtained with current data,
as well as the bounds that could be obtained at
$\sqrt{s}=13$~TeV in the VBF channel for data samples of 100 and 300
fb$^{-1}$.  The CM method has already been successfully applied in the
gluon fusion (GGF) production channel.  The VBF production channel gives important
complementary information, because both production and decay of the
Higgs boson occur already at tree graph level.}

\preprint{FERMILAB-PUB-15-030-T}

\begin{document} 
\maketitle
\flushbottom

\section{Introduction}
\label{sec:intro}

With the advent of 13~TeV running, the LHC will probe in greater
detail the production of the Higgs boson in the sub-dominant mode --
the so-called Vector Boson Fusion (VBF) mechanism. For the observed
Higgs boson of mass $m_H \sim 125$~GeV, the rate in VBF mode is
calculated to be about $8\%$ of the rate in the gluon fusion
mode~\cite{Dittmaier:2011ti} almost independent of the energy of the LHC. 
Because of the increase of energy from
7~TeV to 14~TeV the production cross sections in the gluon fusion
mode and the VBF mode are expected grow by a factor of $\sim 3.3$.

In the context of gluon fusion (GGF) production,
the inadequacy of the narrow width approximation and the importance of
the off-shell tail of the Higgs boson has been first noted by Kauer
and Passarino~\cite{Kauer:2012hd}.  They observed that more than $10\%$ of the Higgs 
cross section $\sigma (gg \to H \to e^-e^+ \mu^- \mu^+)$ lies in the region
where the four-lepton invariant mass, $m_{4l}$ is greater than $130$~GeV. 
The apparent breakdown of the narrow width approximation 
is due to the opening of the threshold for on-shell $Z$-pair production
and the growth with energy of the Higgs boson amplitude.

This point was further developed by Caola and
Melnikov~\cite{Caola:2013yja} who observed that the rates for the
production of the Higgs bosons were given schematically by,
\beqn 
\sigma_{{\rm on-shell}}(gg \to H \to e^-e^+ \mu^- \mu^+)& \sim & \frac{g_i^2 g_f^2}{\Gamma}\; ,
\nonumber \\ 
\sigma_{{\rm off-shell}}(gg \to H \to e^-e^+ \mu^- \mu^+)& \sim & g_i^2 g_f^2 \; ,
\label{eq:CMobservation}
\eeqn 
where $\Gamma$ is the total width of the Higgs boson and
$g_i,g_f$ are the effective couplings of the Higgs boson in the
initial and final state. Thus, observation of the off-shell production
gives access to information about the effective couplings, without
making assumptions on the total width of the Higgs boson.  Moreover, the
ratio of the off-shell and on-shell rates can be used to place
constraints on the total width of the Higgs boson, using accurate theoretical
predictions for the same ratio~\cite{Kauer:2012hd,Caola:2013yja,Campbell:2013una}.  This idea
has been successfully used by the experiments~\cite{Khachatryan:2014iha,ATLAS-CONF-2014-042}, exploiting
primarily the gluon fusion production mechanism.

Turning now to the VBF process, the high mass behaviour of the Higgs
diagram has been the subject of extensive discussion because it gives
access to the longitudinal modes of the vector boson. These polarizations are only 
present because of electroweak symmetry breaking\footnote{For a recent reviews with
extensive references we refer the reader to refs.~\cite{Kilian:2014zja,Szleper:2014ab}.}. 
Indeed the scattering of longitudinal vector bosons was of great importance
in attempts to provide an upper limit on the mass of the Higgs boson~\cite{Lee:1977eg}.
However in the Standard Model it
will be extremely challenging to observe the longitudinal modes in the
next run, because the vector boson polarization
produced by emission from a quark is predominantly transverse~\cite{Dawson:1984gx,Kane:1984bb}.

\begin{figure}[tbp]
\begin{center}
\includegraphics[scale=0.5,angle=270]{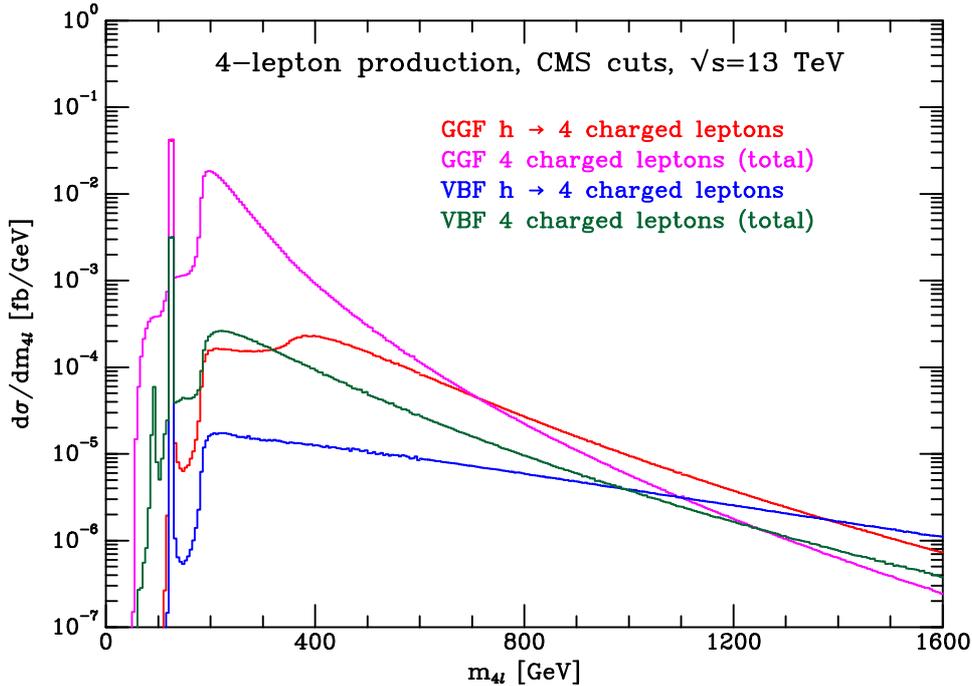}
\end{center}
\caption{\label{comparison} Comparison of the rates for the gluon fusion process
$gg \to H \to e^- e^+ \mu^- \mu^+$ and the vector boson fusion process $qq \to  e^- e^+ \mu^- \mu^+ qq$.
The GGF curves cross at $m_{4l}=700$~GeV, whereas the VBF curves cross at $m_{4l}=1000$~GeV.  
The CMS cuts correspond to those used in Ref.~\protect\cite{Campbell:2013una}
and, for the VBF process, we also apply the cuts specified in
Eqs.~(\ref{cuts:jets}),~(\ref{cuts:vbs}) and~(\ref{cuts:leptbetween}).
The rates are plotted as a function of the four-lepton mass.}
\end{figure}
We can apply the Caola-Melnikov method to bound the Higgs boson width,
using the vector boson fusion process.  As quoted later, current
constraints on the on-shell signal strength place it at about the SM
value. In this case the production and decay of the Higgs boson occurs
at tree level so that this process is sensitive to different
theoretical systematics relative to the gluon fusion process.  In
particular it is not susceptible to loop effects that decouple in the
off-shell region, such as those discussed in
Ref.~\cite{Englert:2014aca}.  Fig.~\ref{comparison} shows a comparison
of the rates for the gluon fusion process and the vector boson fusion process.
As one can see from Fig.~\ref{comparison}, the tail of the Higgs-mediated diagrams
is relatively more important for VBF than for gluon fusion, compared to their respective
peak cross sections.  The differing fall off of the purely Higgs-mediated curves is due to
the growth proportional to $E^2$ ($E$) of the underlying VBF (GGF) amplitudes.

The experimental study of the production of vector boson pairs by the
VBF mechanism is still not fully developed.  Observation of the
related process $p+p \to Z +~{\rm jet}~+~{\rm jet}$ has established
the feasibility of the VBF
method~\cite{Aad:2014dta,Khachatryan:2014dea}.  The results are found
to be in agreement with Standard Model
predictions~\cite{Jager:2012xk,Denner:2014ina}.

Evidence for the $W^\pm W^\pm$ process at $\sqrt{s}=8$~TeV has been presented
by both the ATLAS~\cite{Aad:2014zda} and CMS~\cite{Khachatryan:2014sta}
collaborations. The accumulated luminosity is still too low to permit
stringent VBF cuts on these data samples.  The $W^\pm W^\pm$ process will play
an important role in this paper because of the low backgrounds.  The
$W^-W^+$ channel is the VBF process with the largest rate, although it has
substantial backgrounds from $t\bar t$+jet production.  Note that
although the $W^\pm Z$  and $W^\pm W^\pm$ channels do not contain
diagrams with an $s$-channel Higgs boson, they are sensitive to the
Higgs coupling through $t$-channel exchange diagrams. In this
paper we will use the acronym VBF to denote both vector boson fusion ($s$-channel exchange)
and vector boson scattering ($t$-channel exchange).
The potential to study all vector boson pair channels 
in the next two runs will be discussed in Section~\ref{Basic_rates}.

In the above paragraph we have referred to the VBF processes as  $W^-W^+$,
$W^\pm W^\pm$, $W^\pm Z$ and $ZZ$.
We emphasize that this is only a shorthand for  
all processes that lead to the same 4-lepton final state, i.e.\ all doubly-, singly- 
and non-resonant contributions are included. The doubly resonant processes that define the explicit 
final states are listed below.
\label{WW}
\bothdk{q+q}{W^+ +}{W^-+q+q}{\mu^-+\nu_\mu}{\nu_e+e^+}  
\begin{minipage}[h]{2.95in}
\label{WpWp}
\bothdk{q+q}{W^+ +}{W^++q+q}{\nu_\mu+\mu^+}{\nu_e+e^+} 
\end{minipage}
\begin{minipage}[h]{2.95in}
\label{WmWm}
\bothdk{q+q}{W^- +}{W^-+q+q}{\mu^- + \bar{\nu}_\mu}{e^- + \bar{\nu}_e} 
\end{minipage}
\begin{minipage}[h]{2.95in}
\label{WpZ}
\bothdk{q+q}{W^+ +}{Z/\gamma+q+q}{\mu^-+\mu^+}{\nu_e+e^+} 
\end{minipage}
\begin{minipage}[h]{2.95in}
\label{WmZ}
\bothdk{q+q}{W^- +}{Z/\gamma+q+q}{\mu^-+\mu^+}{e^- +\nu_e}  
\end{minipage}
\begin{minipage}[h]{2.95in}
\label{ZZ}
\bothdk{q+q}{Z/\gamma+}{Z/\gamma+q+q}{\mu^- + \mu^+}{e^- + e^+} 
\end{minipage}
\begin{minipage}[h]{2.95in}
\label{ZZnu}
\bothdk{q+q}{Z/\gamma + }{Z+q+q}{\nu_\mu+\bar{\nu}_\mu}{e^- + e^+}
\end{minipage} \\ \vspace*{0.5cm} \\
In addition to the specific
leptonic final states shown, our phenomenological analysis will 
include other states with
different flavours of leptons $(e,\mu)$ and $(\nu_e,\nu_\mu,\nu_\tau)$. 
Interferences between states 
with identical leptons in the final state are small and will be ignored.

\section{Basic rates for VBF processes}
\label{Basic_rates}
\begin{figure}[tbp]
\begin{center}
\includegraphics[scale=0.7,angle=270]{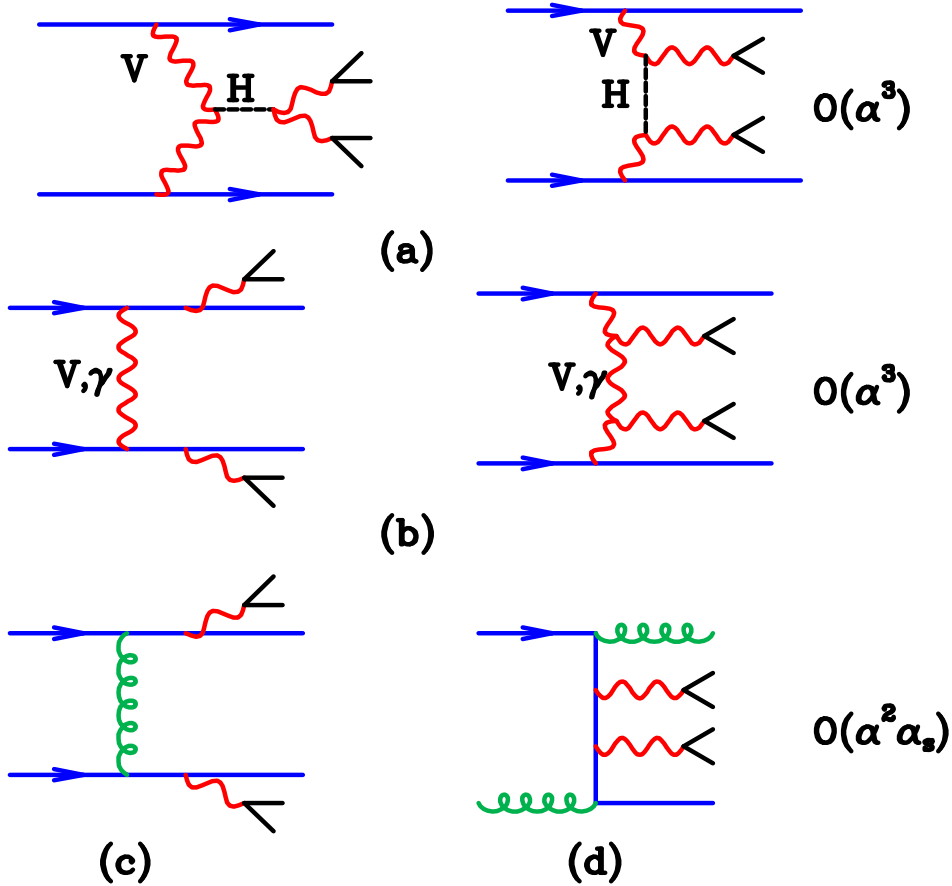}
\end{center}
\caption{\label{diags} Representative Feynman diagrams, where $V$ denotes a $W$ or $Z$ boson.}
\end{figure}

\begin{table}
\begin{center}
\begin{tabular}{ccc}
\hline
\hline
Run 2: & 2015--2017 & $\sim 100~$fb$^{-1}$ \\
Run 3: & 2019--2021 & $\sim 300~$fb$^{-1}$ \\
\hline
\end{tabular}
\caption{Assumed schedule and luminosity of the LHC for the next 7 years.}
\label{Lumi}
\end{center}
\end{table}
We first want to establish which of the VBF modes will be accessible in 
Run 2 and Run 3 of the LHC. The size of the expected data samples is shown in Table~\ref{Lumi}. 
For this exploratory study we will use tree-graph calculations
for which representative diagrams are shown in Figure~\ref{diags}.

The VBF diagrams in which we are primarily interested are shown in
$(a)$ and $(b)$ of the figure and contribute to the amplitude at
$O(\alpha^3)$.  There are also mixed QCD-electroweak diagrams that
lead to the same final state that occur at $O(\alpha^2\alpha_s)$,
shown in diagrams $(c)$ and $(d)$.  These amplitudes have all been
calculated and included in the parton-level integrator,
MCFM~\cite{Campbell:2010ff}.  These tree level amplitudes have
previously been calculated in the program
PHANTOM~\cite{Ballestrero:2007xq,Ballestrero:2010vp} 
and are also available in MadGraph~\cite{Alwall:2014hca}.

In principle we should consider the interference of the electroweak
and QCD-electroweak amplitudes.  However the leading color
interference between quark-quark diagrams $(a,b)$ and $(c)$ vanishes.
The interference, which is color-suppressed, can only occur between a
limited number of processes with identical final state quarks.  In the
phase space region relevant for VBF production, it is very
small~\cite{Ciccolini:2007ec}.  Note that even channels such as $W^\pm
W^\pm$ and $W^\pm Z$, that do not include a $s$-channel Higgs boson
contribution, are still sensitive to the Higgs boson couplings through
$t$-channel exchange.  The tree-level approximation should be
sufficient for a preliminary idea, because the NLO corrections have
been shown to be quite small for these
processes~\cite{Baglio:2014uba,Jager:2011ms,Jager:2013mu}.

\begin{table}
\begin{center}
\begin{tabular}{|c|c|c|c|}
\hline
$m_H$               & 125 GeV                & $\Gamma_H$ & 0.004165 GeV \\
$m_W$               & 80.419 GeV             & $\Gamma_W$ & 2.06 GeV \\
$m_Z$               & 91.188 GeV             & $\Gamma_Z$ & 2.49 GeV \\
$m_t$               & 173.0 GeV              & $m_b$      & 4.75 GeV \\
$e^2$               & 0.0948355              & $g_W^2$    & 0.4267132 \\ 
$\sin^2\theta_W$    & $0.22228 - 0.00131i$   & $G_F$      & $0.116639\times10^{-4}$ \\
\hline
\end{tabular}
\caption{Masses, width and Electroweak parameters used to produce the results in this paper.
\label{parameters}}
\end{center}
\end{table}
Our results have been obtained using the parameters shown in Table~\ref{parameters}, working in the
complex mass scheme where $\sin^2\theta_W$ is related to the complex masses by,
\beq
\sin^2\theta_W = 1-\left(\frac{m_W^2-im_W\Gamma_W}{m_Z^2-im_Z\Gamma_Z}\right)
\eeq
This is important to preserve gauge invariance.
We use the LO pdf set from MSTW2008~\cite{Martin:2009iq} and evaluate all cross-sections
using an event-by-event scale, $\mu$, that is given by the partonic center of mass energy,
$\mu =\sqrt{\hat{s}}$.

In order to select topologies that enhance the vector boson scattering contributions
in comparison with continuum or background contributions, specialized cuts
are required.  
For simplicity, we first apply a set of cuts that is almost the same for all processes.
The two jets are identified by the anti-$k_T$ algorithm,
\begin{equation}
p_{T,J} > 20~{\rm GeV}\,,   |\eta_{J}| < 4.5 \,,   R=0.4 \, .
\label{cuts:jets}
\end{equation}
For the leptons we apply the following cuts:
\begin{eqnarray}
&& p_{T,\ell} > 20~{\rm GeV}\,, \quad |\eta_\ell| < 2.5 \,, \nonumber \\
&& m_{ll} > 10~{\rm GeV}\,, \quad \mbox{for all charged lepton combinations.}
\label{cuts:leptons}
\end{eqnarray}
For processes containing neutrinos the missing transverse energy satisfies,
\begin{equation}
\slsh{E}_T > 40~{\rm GeV}\, .
\label{cuts:met}
\end{equation}

The specialized cuts that isolate the VBF contributions
are as follows.
We ensure that the tagging jets lie in opposite hemispheres,
have a rapidity gap of at least $2.5$ units between them,
and that the invariant mass of the jets, $m_{j_1 j_2}$, is large:
\begin{equation}
y_{gap}  >  2.5, \;\; \eta_1 \times \eta_2  <  0, \;\; m_{j_1 j_2}  >  500~{\rm GeV}.
\label{cuts:vbs}
\end{equation}
As a final cut, we also require that the rapidities of any charged leptons lie between
the two jet rapidities,
\begin{equation}
 \eta_{J}^{min} < \eta_\ell < \eta_{J}^{max}\; .
\label{cuts:leptbetween}
\end{equation}

We now detail the process-dependent additional cut that
has been applied
in each case to isolate the off-shell region.  For the four charged-lepton process
we cut on $m_{4l}$, the invariant mass of the four lepton system, 
\beq \label{mcut4l}
m_{4l}^2 =(p_{l_1} +p_{l_2} +p_{l_3} +p_{l_4} )^2\, ,
\eeq
while for the other $ZZ$ process we use the transverse mass $m_T^{ZZ}$ defined by,
\beq \label{mcutzz}
(m_T^{ZZ})^2 = \left[ \sqrt{m_Z^2 + p_{T, ll}^2} + \sqrt{m_Z^2 + E_{T, miss}^2} \right]^2
 - \left| \vec{p}_{T, ll} + \vec{E}_{T, miss} \right|^2 \;.
\eeq
For the $WW$ processes we define the transverse mass as,
\beq \label{mcutww}
(m_T^{WW})^2 = \left[ E_{T, ll} + E_{T, miss} \right]^2
 - \left| \vec{p}_{T, ll} + \vec{E}_{T, miss} \right|^2 \;,
\eeq
where $E_{T,ll} = \sqrt{p_{T,ll}^2 + m_{ll}^2}$.
The appropriate transverse
variable for the $WZ$ processes is,
\beq \label{mcutwz}
(m_T^{WZ})^2 = \left[ \sqrt{m_{3l}^2 + p_{T, 3l}^2} + E_{T, miss} \right]^2
 - \left| \vec{p}_{T, 3l} + \vec{E}_{T, miss} \right|^2 \;.
\eeq
For the two-lepton processes, $W^-W^+$, $W^\pm W^\pm$ and $ZZ\to 2l2\nu$, we
impose additional cuts in order to reduce the backgrounds from top quark
processes.  We require that two combinations of lepton and jet invariant
masses are above the top mass~\cite{Szleper:2014ab},
\begin{equation}
m_{l_1 j_2}, m_{l_2 j_1}  > 200~{\rm GeV}  \;\; .
\label{cuts:mlj}
\end{equation}
For these cuts $l_1$, $l_2$ ($j_1$, $j_2$) represent the leptons (jets) of
highest and lowest transverse momentum respectively.

The processes that we will study in this paper, together with the cross sections after imposing the above set of cuts,
are shown in Tables~\ref{weakxsecs} (electroweak process) and~\ref{weakqcdxsecs} (mixed QCD-electroweak).
We also show the expected number of events in $100$~fb$^{-1}$ of data, after all flavors of lepton have
been included.  Note that, in this paper, we have consistently neglected $\tau$-leptons.  In principle, including
$\tau$-leptons would increase the anticipated event rates,  by a factor of $9/4$ for channels such as $W^- W^+$,
but in practice the increase would not be so large because of the limited efficiency for $\tau$-identification.
From the results in these tables it is clear that the final states of highest interest in the immediate future
are the two charged-lepton ones. Despite being relatively background-free, the $ZZ\to 4l$ rate is so small that
the SM expectation is for only a few events.  Similarly the $ZZ \to 2l2\nu$ channel with a $m_T^{ZZ}$ cut
anticipating $Z$-pair events produces only $11$ events and suffers from much larger backgrounds.
The $W^-W^+ \to 2l2\nu$ channel has a very high rate and in reality cannot be separated from the 
$ZZ \to 2l2\nu$ channel.  For this reason we consider both processes with a $m_T^{WW}$ cut; the event
rates will be added in subsequent sections.  The same-sign dilepton processes $W^\pm W^\pm \to 2l2\nu$
are, in total, only about a factor of three smaller than the opposite-sign one.
The mixed QCD-electroweak processes in Table~\ref{weakqcdxsecs} represent significant, irreducible backgrounds
which in most cases are of similar size as the electroweak processes.  
These backgrounds are reduced to this level by the imposition of the VBF cuts.
The notable exceptions are once again same-sign 
$W^+W^+$ and $W^-W^-$ production, where the pure electroweak processes are larger by an order of magnitude.

\begin{table}
\begin{center}
\begin{tabular}{|l|l|c|c|c|c|}
\hline
Process                                    & Nominal                  & Cut                     &  $\sigma$ [fb] & Factor & Events\\
                                           & process                  &                         &  $O(\alpha^6)$ &        & in 100 fb$^{-1}$ \\
\hline
$pp \to e^-\mu^+ \nu_\mu \bar{\nu}_e jj$   & $W^-W^+$                & $m_T^{WW} >300$~GeV      & 0.2378   &x4 & 95 \\
\hline
$pp \to \nu_e e^+ \nu_\mu \mu^+ jj$        & $W^+W^+$                & $m_T^{WW} >300$~GeV      & 0.1358   &x2 & 27 \\
$pp \to e^-\bar{\nu_e}\mu^-\bar{\nu_\mu}jj$& $W^-W^-$                & $m_T^{WW} >300$~GeV      & 0.0440   &x2 & 9 \\
\hline
$pp \to \nu_e e^+ \mu^- \mu^+ \mu^+ jj$    & $W^+Z$                  & $m_T^{WZ} >300$~GeV      & 0.0492   &x4 & 20 \\
$pp \to e^-\bar{\nu_e} \mu^-\mu^+ jj$      & $W^-Z$                  & $m_T^{WZ} >300$~GeV      & 0.0242   &x4 & 10 \\
\hline
$pp \to l^-l^+\nu_l \bar{\nu}_ljj$         & $ZZ$                    & $m_T^{ZZ} >300$~GeV      & 0.0225   &x6 & 14 \\
$pp \to l^-l^+\nu_l \bar{\nu}_ljj$         & $ZZ$                    & $m_T^{WW} >300$~GeV      & 0.0181   &x6 & 11 \\
$pp \to e^-e^+\mu^-\mu^+jj$                & $ZZ$                    & $m_{4l} >300$~GeV        & 0.0218   &x2 & 4 \\
\hline
\end{tabular}
\caption{Electroweak (${\cal O}(\alpha^6)$) cross sections at $\sqrt{s}=13$~TeV, under the cuts
given in Eqs.~(\protect\ref{cuts:jets})--(\protect\ref{cuts:leptbetween}) and the off-shell definition specified
in the table. The factor gives the approximate number by which the result shown for specific lepton flavours 
must be multiplied to account for two flavours of charged leptons, $e,\mu$ and three flavours of neutral leptons, $\nu_e, \nu_\mu, \nu_\tau$.
\label{weakxsecs}}
\end{center}
\end{table}

\begin{table}
\begin{center}
\begin{tabular}{|l|l|c|c|c|c|}
\hline
Process                                    & Nominal                  & Cut                     &  $\sigma$ [fb]           & Factor & Events\\
                                           & process                  &                         &  $O(\alpha^4\alpha_s^2)$ &        & in 100 fb$^{-1}$ \\
\hline
$pp \to e^-\mu^+ \nu_\mu \bar{\nu}_e jj$   & $W^-W^+$                & $m_T^{WW} >300$~GeV      & 0.2227   &x4 & 89 \\
\hline
$pp \to \nu_e e^+ \nu_\mu \mu^+ jj$        & $W^+W^+$                & $m_T^{WW} >300$~GeV      & 0.0079   &x2 & 2 \\
$pp \to e^-\bar{\nu_e}\mu^-\bar{\nu_\mu}jj$& $W^-W^-$                & $m_T^{WW} >300$~GeV      & 0.0025   &x2 & 0 \\
\hline
$pp \to \nu_e e^+ \mu^- \mu^+ \mu^+ jj$    & $W^+Z$                  & $m_T^{WZ} >300$~GeV      & 0.0916   &x4 & 37 \\
$pp \to e^-\bar{\nu_e} \mu^-\mu^+ jj$      & $W^-Z$                  & $m_T^{WZ} >300$~GeV      & 0.0454   &x4 & 18 \\
\hline
$pp \to l^-l^+\nu_l \bar{\nu}_ljj$         & $ZZ$                    & $m_T^{ZZ} >300$~GeV      & 0.0143   &x6 & 9 \\
$pp \to l^-l^+\nu_l \bar{\nu}_ljj$         & $ZZ$                    & $m_T^{WW} >300$~GeV      & 0.0118   &x6 & 7 \\
$pp \to e^-e^+\mu^-\mu^+jj$                & $ZZ$                    & $m_{4l} >300$~GeV        & 0.0147   &x2 & 3 \\
\hline
\end{tabular}
\caption{Mixed QCD-electroweak (${\cal O}(\alpha^4\alpha_s^2)$) cross sections at $\sqrt{s}=13$~TeV, under the cuts
given in Eqs.~(\protect\ref{cuts:jets})--(\protect\ref{cuts:leptbetween}) and the off-shell definition specified
in the table.
\label{weakqcdxsecs}}
\end{center}
\end{table}

The same-sign lepton channels are relatively
background-free\footnote{The double parton scattering contribution is
negligible with our cuts~\cite{Gaunt:2010pi}.}  and the $ZZ \to 4l$
channel suffers only small backgrounds. However the processes
involving neutrinos are potentially subject to large backgrounds from
top production, possibly in association with additional jets.  Further
sources of background include $W+$~jets events and QCD multijets, where
jets are misidentified as leptons.  We do not include a study of the
effects of such misidentification in this paper.  To assess the
importance of top quark processes we have computed the LO
cross-section for $t\bar t$ production under the same set of cuts.  We
find, in the off-shell tail defined by $m_T^{WW} >300$~GeV,
\begin{equation}
\sigma(t\bar t \to e^-\bar\nu_e b \mu^+\nu_\mu \bar b)=0.637~\mbox{fb} \;,
\end{equation}
which corresponds to 254 total events, summing over lepton flavors, in
100 fb$^{-1}$ of data.  Although this estimate ignores both the effect
of higher orders, and the application of a $b$-jet veto to suppress
some of these events, it is sufficient to illustrate the severity of
this background.  This background rate is larger than the
expected signal for $W^-W^+$ production given in
Table~\ref{weakxsecs}.  Even if the background can be somewhat reduced
by application of a $b$-jet veto or other cuts, it will still present
a significant background with attendant uncertainties.  For the
same-sign channels this background is relatively insignificant since
it would only enter if the charge of one of the leptons were
misidentified, which typically occurs at the sub-percent
level~\cite{Szleper:2014ab}.

Given the size of the event rates in Table~\ref{weakxsecs} and the
background levels discussed above, it is clear that the most important
process for Run 2 is like-sign production, $W^\pm W^\pm$. However 
in Run 3, some of the other processes will also contribute useful information.
In the following section we will present results for all
processes for the sake of comparison.
In addition, with sufficient understanding
of top quark backgrounds, it may eventually be possible to extricate
the opposite-sign $W^-W^+$ channel from the top background. At present
we are considering analyses based on cuts only. With sufficient data
one can move to more sophisticated analyses, e.g.\ ones based on
multivariate methods, that take more detailed
information about the matrix elements into account.

\section{Sensitivity to couplings and the Higgs boson width}
We now turn to the matter of the sensitivity of the VBF cross sections to the Higgs boson couplings.
We will work in the interim framework for the analysis of  Higgs couplings~\cite{Heinemeyer:2013tqa} where
the couplings of the Higgs to $W$ and $Z$ bosons scale in the same way,
\beq
\frac{\Gamma_{WW}}{\Gamma^{SM}_{WW}}  =  \kV^2,\;\;\;
\frac{\Gamma_{ZZ}}{\Gamma^{SM}_{ZZ}}  =  \kV^2 \; .
\label{eq:couplingrescaling}
\eeq 
As indicated before, the advantage of the off-shell measurements
is that they give information about the Higgs couplings, without
assumptions on the total width of the Higgs boson.  Although the
parameter range $ \kV < 1$ is perhaps better motivated from a
theoretical point of view, we shall consider both $\kV < 1$ and $\kV >
1$. The latter range comes into play when we try and constrain the
total Higgs width using the CM method~\cite{Caola:2013yja}. Within the
standard model, and after the Higgs boson discovery, the Higgs
couplings are completely determined. The interim framework attempts to
capture BSM features in an approximate way.

As more data is accumulated, the emphasis will move from this interim
framework of effective couplings, to a more general approach using higher dimension
operators in an effective theory.  The effective field theory operators parameterize
the observed behaviour at current energies in terms of towers of
higher dimension operators, normally chosen to be CP-invariant and
invariant under $SU(2) \times U(1)$ local symmetry.  The effective operators
are the consequence of unknown physics occurring at a higher scale
$\Lambda$. By assumption the scale $\Lambda$ is higher than the scale
currently being probed and the importance of the operators is
controlled by their dimension.

A complete operator analysis is beyond the scope of this paper.
We shall consider one particular operator of dimension six in order 
to make a connection with our interim framework,
\beq \label{HD}
  \LL_{HD} = F_{HD}\ 
  \tr{{\vH^\dagger\vH}- \frac{v^2}{4}}\cdot
  \tr{\left (\vD_\mu \vH \right )^\dagger \left (\vD^\mu \vH \right )} \; .
\eeq
where $\vD = \partial_\mu  - i  g \vW_\mu  - i g^\prime \vB_\mu$ is the
$SU(2)\otimes U(1)$ covariant derivative.
$\LL_{HD}$ leads to the following Feynman rules for the couplings of a single Higgs boson
in the unitary gauge,
\beqn
  hW^+_\mu W^-_\nu:& \hspace{0.5cm} &  i g M_W  g_{\mu\nu} \frac{v^2 F_{HD}}{2}  \;,
\\
  hZ_\mu Z_\nu:& \hspace{0.5 cm} &
  i g \frac{M_W}{\cos^2\theta_W}  g_{\mu\nu} \frac{v^2 F_{HD}}{2}  \;.
\eeqn
Operators of dimension six that contribute
to triple gauge boson couplings (TGC)
can be ignored in the present context, since their coefficients are
exquisitely constrained by TGC measurements~\cite{Corbett:2013pja}.
The operator in Eq.~(\ref{HD}), if we ignore unitarization effects, 
just corresponds to an effective coupling.
\beq
\kV=1+F_{HD} \frac{v^2}{2} \;,
\eeq
where $v=0.246$~TeV and $F_{HD} $ can have either sign. 
Note that useful reference points are obtained by choosing $F_{HD}=\pm 30$~TeV$^{-2}$,
which correspond to $\kV=2$ and $\kV=0$ respectively. 
However the validity of the effective field theory 
is seriously compromised for a value of $F_{HD}$ that large, since the corresponding scale $\Lambda$ is too low.
Conversely, to probe scales higher than $1$~TeV will nominally require sensitivity to $\kV$ at the $3\%$ level.
 
The rescaling in Eq.~(\ref{eq:couplingrescaling}) implies that the off-shell cross section in the VBF final
states should be different from the Standard Model.  Due to the interference between diagrams that involve the
Higgs boson and those that do not, the off-shell cross sections do not simply grow with $\kV^4$, as suggested
by Eq.~(\ref{eq:CMobservation}).  Instead there is a term proportional to $\kV^2$ resulting from the interference,
whose coefficient is negative because the interference is destructive.
Explicitly, we find the number of off-shell events in $100$~fb$^{-1}$, due to the electroweak process only,
is given by,
\begin{eqnarray} \label{kappacurves}
l^- l^+ \nu\bar\nu :      &&  \quad N^{\off} =  127.9 - 42.8 \, \kV^2 + 20.8 \, \kV^4  \nonumber \\
l^+ l^+ \nu\nu  :         &&  \quad N^{\off} =  37.2  - 18.3 \, \kV^2 + 8.3  \, \kV^4  \nonumber \\
l^- l^- \bar\nu\bar\nu  : &&  \quad N^{\off} =  11.0  - 4.1  \, \kV^2 + 1.8  \, \kV^4  \nonumber \\
l^+ l^- l^+ \nu  :        &&  \quad N^{\off} =  23.5  - 6.8  \, \kV^2 + 3.2  \, \kV^4  \nonumber \\
l^+ l^- l^-\bar\nu  :     &&  \quad N^{\off} =  11.3  - 3.3  \, \kV^2 + 1.6  \, \kV^4  \nonumber \\
l^- l^+ l^- l^+ :         &&  \quad N^{\off} =   6.0  - 3.0  \, \kV^2 + 1.5  \, \kV^4  
\end{eqnarray}
where, for the $l^- l^+ \nu \bar \nu$ channel we have added the $W^- W^+$ and $ZZ$ results for $m_T^{WW}>300$~GeV
from earlier, with the appropriate multiplicative factors.
Adding in the irreducible background from the mixed QCD-electroweak processes in Table~\ref{weakqcdxsecs} we obtain,
\begin{eqnarray} \label{kappamixedcurves}
l^- l^+ \nu\bar\nu :      &&  \quad N^{\off} =  224.8 - 42.8 \, \kV^2 + 20.8 \, \kV^4  \nonumber \\
l^+ l^+ \nu\nu  :         &&  \quad N^{\off} =  38.8  - 18.3 \, \kV^2 + 8.3  \, \kV^4  \nonumber \\
l^- l^- \bar\nu\bar\nu  : &&  \quad N^{\off} =  11.5  - 4.1  \, \kV^2 + 1.8  \, \kV^4  \nonumber \\
l^+ l^- l^+ \nu  :        &&  \quad N^{\off} =  60.1  - 6.8  \, \kV^2 + 3.2  \, \kV^4  \nonumber \\
l^+ l^- l^-\bar\nu  :     &&  \quad N^{\off} =  29.5  - 3.3  \, \kV^2 + 1.6  \, \kV^4  \nonumber \\
l^- l^+ l^- l^+ :         &&  \quad N^{\off} =   9.0  - 3.0  \, \kV^2 + 1.5  \, \kV^4  
\end{eqnarray}
Note that the numbers of events are not so different for $\kV=0$, (no
Higgs boson) and for $\kV=1$, (SM). This reflects the fact that, for
this energy and luminosity, we cannot place the off-shell mass cut at
a value sufficiently far above the electroweak scale, $v=246$~GeV that
the terms that fail to cancel for $\kV \neq 1$ dominate.

\subsection{Limit from Run 1}

The existing analyses of $W^\pm W^\pm$ production in $\sqrt{s}=8$~TeV running by
both the ATLAS~\cite{Aad:2014zda} and CMS~\cite{Khachatryan:2014sta}
collaborations already provide a constraint on the Higgs boson couplings
and width.  Since the luminosity accumulated in Run 1 is too low to clearly isolate
the VBF region, the expected sensitivity is rather limited.

As an example, we have repeated our analysis using the cuts presented in
Ref.~\cite{Aad:2014zda}.  We find that the cross-section for two electrons
(or positrons), computed in the fiducial region defined there, is:
\beq
\sigma^{same-sign}_{electron} =  0.345 - 0.036  \, \kV^2 + 0.014 \, \kV^4 \, \mbox{fb}\;.
\eeq 
This does not yet account for multiple lepton flavors or efficiencies.   We
therefore normalize to the SM expectation quoted by ATLAS,
$\sigma = 0.95 \pm 0.06$~fb, to obtain,
\beq
\sigma^{same-sign}_{fiducial} =  1.015 - 0.106  \, \kV^2 + 0.040 \, \kV^4 \, \mbox{fb}\;.
\eeq
The fiducial-region measurement quoted by ATLAS is,
\beq
\sigma^{measured} =  1.3 \pm 0.4 (stat) \pm 0.2 (syst) \, \mbox{fb}\;.
\eeq
Accounting for a $2\sigma$ deviation in the sum of statistical, systematic and theoretical uncertainties
we thus obtain the coupling constraint,
\beq
\kV < 7.8 \,. 
\eeq
We now turn to the matter of the sensitivity of the VBF cross sections to the Higgs boson width, under the
assumption that all Higgs boson couplings and widths scale in a manner that leaves the on-shell cross sections
unchanged.  This assumption is supported by the latest analyses of Run 1 LHC data from ATLAS and CMS that constrain
the on-shell signal strength $\mu = \sigma^{observed}/\sigma^{SM}$. By analysing the $H \to WW$ decay channel,
the ATLAS experiment obtains the following bound~\cite{ATLAS:2014aga},
\beq
\mu_{VBF}^{ATLAS} = 1.27^{+0.53}_{-0.45} \;,
\eeq
while the CMS experiment finds~\cite{Khachatryan:2014jba},
\beq
\mu_{VBF}^{CMS} = 1.16^{+0.37}_{-0.34} \;.
\eeq
In order for the on-shell signal strength $\mu$ to be equal to unity, this implies that the width of the
Higgs boson should scale as,
\beq
\Gamma_H \to \kV^4 \Gamma_H^{SM} \;.
\eeq
Making the simplifying assumption that the signal strength is exactly equal to one
with no error, we obtain an upper-bound on the Higgs boson width of,
\beq
\Gamma_H < 60.8 \times \Gamma_H^{SM} \;.
\eeq
This simple analysis has many shortcomings; it is a leading order analysis
without matching experimental cuts for the on-shell $W^-W^+$ and the off-shell
$W^\pm W^\pm$ samples.
We present it only to make the basic point that off-shell couplings may be best
constrained by the $W^\pm W^\pm$ channels, despite the fact they do not contain 
the $s$-channel exchange of a Higgs boson.

\subsection{Limits from Runs 2 and 3}
To assess the utility of the VBF processes in the immediate run, we
first perform a simple analysis to determine the optimal cut to isolate
the off-shell tail.  In this analysis we obtain the SM prediction for
the number of events ($N$) for all values of the off-shell $m_{cut}$,
where the cut variable depends on the process under consideration.
For the 4-charged lepton process we require $m_{4l} > m_{cut}$,
Eq.~(\ref{mcut4l}), and for the other processes we use
Eqs.~(\ref{mcutzz},\ref{mcutww},\ref{mcutwz}) as appropriate. The SM
expectation includes both the electroweak and the mixed
QCD-electroweak processes, although not the (small effects of)
interference between these two types of process.  Defining a purely
statistical uncertainty by $\delta_N = \sqrt{N}$, we consider which
values of $\kV$ could be excluded at approximately $95 \%$ c.l.\ by an
observation of $N+2\delta_N$ events.  Since the lower bound on
$\kV$ is most useful, we then choose the value of $m_{cut}$ that
provides the strongest such bound, $\kV > \kV^{min}$.  If no lower
bound is obtained we optimize with respect to the upper bound.  In all
cases, we ensure that the value of $m_{cut}$ corresponds to a SM
prediction of at least 10 events.  This process is repeated for
samples of both $100$ and $300$~fb$^{-1}$ of data.

\begin{figure}[tbp]
\begin{center}
\includegraphics[scale=0.4,angle=270]{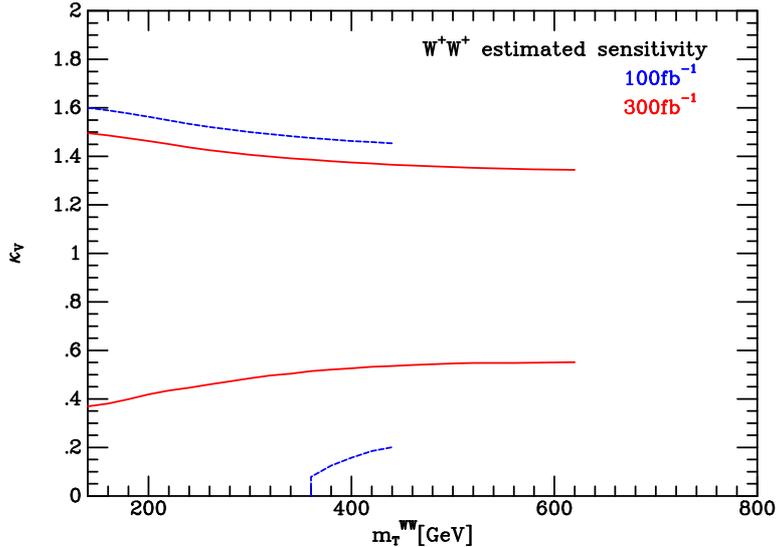}
\end{center}
\caption{\label{fig:senswpwp}
The upper and lower bounds on $\kV$ obtained from
$W^+W^+$ events, as a function of the cut on the
transverse mass, $m_T^{WW}$. The bounds are obtained as described in the 
text. Limits from $100$~fb$^{-1}$ of data
are shown as dashed (blue) lines and those from $300$~fb$^{-1}$
are indicated by solid (red) lines.}
\end{figure}

With this simple procedure we find that, with $100$~fb$^{-1}$ of data,
only the $W^+W^+$ process provides a lower bound.
The limits are shown in Figure~\ref{fig:senswpwp}.
This data set is statistically limited and we find that the best limit
corresponds to the largest cut allowed before the SM expectation falls
below 10 events.  At this value of the cut, $m_{cut} = 440$~GeV, we
find,
\begin{equation}
0.20 < \kV < 1.45 \;.
\end{equation}
The upper bounds obtained from the other processes are,
\begin{equation}
\kV < (1.47, 1.72, 1.66, 1.83, 1.75) \;,
\end{equation}
for, respectively, the processes $(W^-W^+, W^-W^-, W^+Z, W^-Z, ZZ)$.
Figure~\ref{fig:senswpwp} also shows the corresponding bounds expected
with $300$~fb$^{-1}$ of data, where a similar pattern is observed.  The
best lower limit comes again from the $W^+W^+$ process
and corresponds to saturating the 10-event minimum,
which is now reached with $m_{cut}=620$~GeV, and we find,
\begin{equation}
0.55 < \kV < 1.34 \;.
\end{equation}

\begin{figure}[tbp]
\begin{center}
\includegraphics[scale=0.4,angle=270]{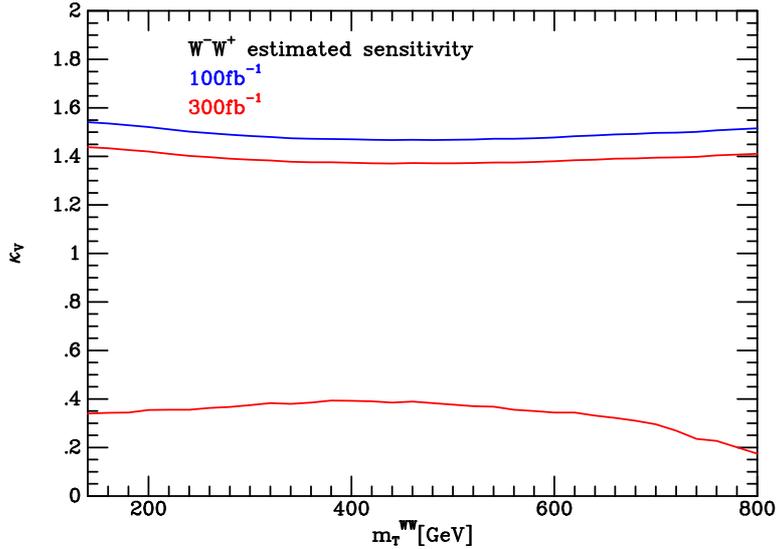}
\end{center}
\caption{\label{fig:senswmwp}
The upper and lower bounds on $\kV$ obtained from
$W^-W^+$ events,
as a function of the cut on the
transverse mass, $m_T^{WW}$ . Limits from $100$~fb$^{-1}$ of data
are shown as dashed (blue) lines and those from $100$~fb$^{-1}$
are indicated by solid (red) lines.}
\end{figure}

With a bigger data set (300~fb$^{-1}$) we can also obtain lower bounds from the
$W^-W^+$ process.  The sensitivity is indicated in
Figure~\ref{fig:senswmwp} and we find,
\begin{eqnarray}
0.39 < \kV < 1.38 \,, && \quad W^-W^+ \; (m_T^{WW}>380~\mbox{GeV}) \;.
\end{eqnarray}
The optimum for the $W^-W^+$ channel displays a real trade-off
as $m_{cut}$ is increased, between decreasing statistics and increasing sensitivity.
Upper limits on $\kV$ from the other processes are weaker.

\begin{figure}[tbp]
\begin{center}
\includegraphics[scale=0.4,angle=270]{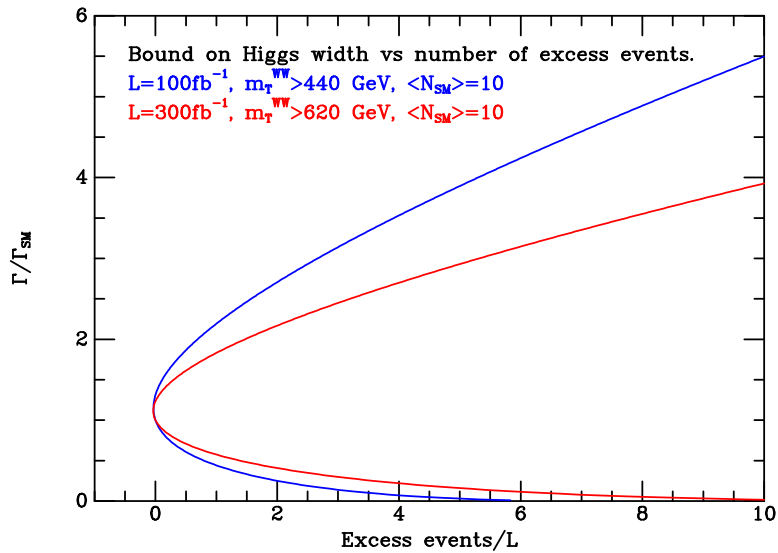}
\end{center}
\caption{\label{excess} 
Bound on the Higgs width vs number of excess off-shell events in $100$ and $300$~fb$^{-1}$}
\end{figure}

We now turn to the matter of the sensitivity of the VBF cross sections to the Higgs boson width,
under the same assumptions as discussed previously, c.f.\ Section 3.1.
The best upper limits on $\kV$ are obtained from the $W^+W^+$ process.  Converting
these into an expected bound on the width from this channel we obtain,
\begin{eqnarray}
\Gamma_H < 4.4 \times \Gamma_H^{SM} \qquad (100~\mbox{fb}^{-1}~\mbox{data}) \;, \nonumber \\
\Gamma_H < 3.2 \times \Gamma_H^{SM} \qquad (300~\mbox{fb}^{-1}~\mbox{data}) \;.
\end{eqnarray}
Fig.~\ref{excess} shows the dependence of the limit on the number of excess events for
100 fb$^{-1}$ and 300 fb$^{-1}$, in the ranges $m_T^{WW}>440$~GeV
and $m_T^{WW}>620$~GeV respectively.
These bounds roughly correspond to the current constraints from
the gluon-fusion channel obtained in Run 1 of the LHC~\cite{Khachatryan:2014iha,ATLAS-CONF-2014-042}.
Previous estimates of the bounds possible with even higher luminosity are given in ref.~\cite{Englert:2014aca}.

\section{Conclusions}
We refocus attention on the fact that in Runs 2 and 3 at the LHC there will be sensitivity
to the electroweak production of four lepton final states in the VBF mode. 
The most important channel
will be $W^+ W^+$, which despite the absence of an $s$-channel Higgs pole, still has sensitivity to
the Higgs coupling. 

As emphasized by Caola and Melnikov~\cite{Caola:2013yja}, the high
mass tail in these final states will give access to information about
the couplings of the Higgs bosons which is free from assumptions about
the total width of the Higgs boson. In the first instance this
information will be analysed in the effective coupling approximation
which parameterizes the strength of the couplings to vector bosons as
a multiple $\kV$ of their standard model strength. 
Because of interference with non-Higgs mediated contributions, the
sensitivity to $\kV$ decreases as the Standard model value $\kV=1$ is approached.
We find that for the event samples likely to be accumulated in the next decade,
the constraints on the effective couplings are quite modest, and consequently a full effective operator
analysis of the constraints from these channels is probably premature.

Conversely, if we assume that the off-shell Higgs couplings are the
same as the on-shell Higgs couplings, we can use the size of the
off-shell contributions to place bounds on the total width of the
Higgs boson. We have pointed out that the off-shell contributions can also be
measured in channels such as $W^\pm W^\pm, W^\pm Z$ where there is no
Higgs boson peak. These have smaller backgrounds than the
dominant $W^- W^+$ channel. Although, as we have seen, the most stringent single channel is 
$W^\pm W^\pm$, results from all channels could be combined in the effective coupling framework.

\acknowledgments

This research is supported by the US DOE under contract DE-AC02-07CH11359.
We are happy to acknowlege useful discussions with Estia Eichten and Chris Quigg.

\bibliographystyle{JHEP}
\bibliography{ZZ}

\end{document}